\documentclass[aps,twocolumn,pra,showpacs]{revtex4}
\usepackage{graphicx}
\usepackage{amsmath,amssymb,amsthm,bbm,latexsym}
\usepackage{amsthm}
\usepackage{amsbsy}

\begin{document}

\title{Remote controlled-NOT gate of $d$-dimension}

\begin{abstract}
Single qubit rotation gate and the controlled-NOT (CNOT) gate constitute a
complete set of gates for universal quantum computation. In general the CNOT
gate is only for two nearby qubits. For two qubits which are remote from
each other, we need a series of swap gates to transfer these two qubits to
the nearest neighboring sites, and then after the CNOT gate we should
transfer them to their original sites again. However, a series of swap gates
are resource for quantum information processing. One economy way which does
not consume so much resource is to implement CNOT gate remotely. The remote
CNOT gate is to implement the CNOT gate for two remotely separated qubits
with the help of one additional maximally entangled state. The original
remote CNOT gate is for two qubits, here we will present the $d$-dimensional
remote CNOT gate. The role of quantum teleportation is identified in the
process of the remote CNOT gate.
\end{abstract}

\author{Gui-Fang Dang and Heng Fan}
\affiliation{Institute of Physics, Chinese Academy of Sciences, Beijing 100080, China}
\maketitle

Quantum information and quantum computation have been attracting a great
deal of interests in the past years since some computing tasks can be sped
up on a quantum computer \cite{CNOT}.

To perform the universal quantum computation, we only need several
elementary gates to be implemented \cite{BBCD}. For example, the single
qubit rotation gate and the controlled-NOT (CNOT) gate constitute a complete
set of gates for universal quantum computation. The CNOT gate is a two-qubit
gate which involves a global unitary transformation on two qubits
coherently, and it can not be realized individually on two qubits
separately. This is because that actually CNOT gate can create a maximally
entangled state from a product (thus separate) state and it is nonlocal\cite%
{remote CNOT}. In real quantum systems, the qubits only interact with each
other when they are in nearest neighboring sites. It is not free to move one
qubit from one site to another site, which means the quantum state transfer
is not free in real quantum systems. Compared with a quantum system where
the quantum state can be transferred freely, extra error penalty should be
paid for the fault tolerant quantum computation with only nearest
neighboring communication \cite{SBF}. Thus to implement a CNOT gate on two
qubits, we should assume that these two qubits are in nearest neighboring
sites so that they can interact with each other and in this way the CNOT
gate realization is possible. To move one qubit from one-site to another, we
need a series of swap gates, each of them swap two nearest neighboring
qubits. The swap gate can be realized by several CNOT and single-qubit
rotation gates. However, this scheme can be considered to be resource
consuming since we generally want the quantum circuit involving only a small
amount of quantum gates. One way to solve this problem is to transfer
quantum state through a quantum bus as demonstrated recently in experiments
\cite{bus1,bus2}. In this paper, we will consider another way which is to
construct a remote CNOT gate with the expense of a prior maximally entangled
state \cite{H Representation,gottesman,DK,remote CNOT,remote CNOT1,remote
CNOT2,remote CNOT3,remote CNOT4}. The advantages of the remote CNOT gate
are, for example: the remote CNOT involves only a few quantum gates and thus
the error rate will be low; we only need to concern about how well the
quality of the shared entangled state is, and we do not need to care too
much the decoherence in the chain of qubits. While the remote CNOT gate is
only for two-qubit system, the $d$-dimensional remote CNOT gate is still
absent. In this paper, we will present the form of the general remote CNOT
gate in $d$-dimensional quantum system. And the role of quantum
teleportation is identified explicitly in this remote CNOT gate.

One of the most amazing protocols in quantum information processing is the
well-known quantum teleportation \cite{teleportation}. In quantum
teleportation, a quantum state can be transferred from one place to another
place by only local operations and classical communication when a prior
maximally entangled state is shared between these two spatially separated
parties. It is not only useful in quantum information transfer, but also
primitive in the universal quantum computation, for example, if the quantum
teleportation can be realized, then the universal quantum computation can be
achieved with only measurements \cite{GC}.

In the quantum teleportation demonstrated in Ref.\cite{teleportation}, to
teleport the unknown state of qubit $A^{\prime }$, Alice needs to perform a
Bell measurement on qubits $A$ and $A^{\prime }$. Then Alice tells Bob the
result of her measurement through classical channel. Bob takes corresponding
operation on qubit $B$ to obtain the state of $\left\vert \phi \right\rangle
_{B}$. We would like to remark that so far, Bell measurement is not easy to
be performed experimentally \cite{Bell measurement1,Bell measurement2,Bell
measurement3}. However, with CNOT gate available, the teleportation without
Bell measurement can be performed \cite{H Representation}. Unlike Bell
measurement, Alice only needs to measure the two individual qubits in their
computational basis respectively. In this scheme, CNOT play an nontrivial
role because CNOT can change entangled states into product states. On the
other hand, as is well known, it also has the entanglement power.

In this paper, we will firstly show that the $d$-dimensional CNOT gate plays
a same role in the $d$-dimensional single-qudit measurement quantum
teleportation (hereafter, the $d$-dimensional state is named as qudit to
correspond the qubit for two-dimensional quantum state), which is helpful to
understand its application in remote CNOT of $d$-dimension. Then, by
generalizing the remote CNOT in two-dimension to higher dimension, we
present the remote CNOT in $d$-dimensional Hilbert space. And the relation
between teleportation and remote CNOT are discussed.

To begin with, let's introduce some notations. In $d$-dimensional Hilbert
space, $U_{mn}=X^{m}Z^{n}$ $\left( m,n=0,1\ldots ,d-1\right) $ are
generalized Pauli matrices constituting a basis of unitary operators, and $%
X\left\vert j\right\rangle =\left\vert \left( j+1\right) \text{mod}%
d\right\rangle $, $Z\left\vert j\right\rangle =e^{i2\pi j/d}\left\vert
j\right\rangle $, where $\left\{ \left\vert j\right\rangle \right\}
_{j=0}^{d-1}$ is an orthonormal basis. The generalized CNOT operator and
Hadamard operator are $C_{01}=\overset{d-1}{\underset{j,k=0}{\sum }}%
\left\vert j\right\rangle _{0}\left\langle j\right\vert \otimes \left\vert
k+j\right\rangle _{1}\left\langle k\right\vert $ and $H=\frac{1}{\sqrt{d}}%
\overset{d-1}{\underset{j,k=0}{\sum }}e^{i2\pi jk/d}\left\vert
k\right\rangle \left\langle j\right\vert $. State at site $0$ is the
control-qudit, state at site 1 is the target-qudit. Arbitrary single-qudit
state in $d$-dimensional Hilbert space can be expressed as%
\begin{equation}
\left\vert \phi \right\rangle _{0}=\overset{d-1}{\underset{j=0}{\sum }}%
a_{j}\left\vert j\right\rangle _{0},  \label{1}
\end{equation}%
where $\overset{d-1}{\underset{j=0}{\sum }}\left\vert a_{j}\right\vert ^{2}=1
$. Let Alice and Bob share an maximally entangled state%
\begin{equation}
\left\vert \Psi ^{\left( 0\right) }\right\rangle _{12}=\frac{1}{\sqrt{d}}%
\overset{d-1}{\underset{j=0}{\sum }}\left\vert jj\right\rangle _{12}.
\label{2}
\end{equation}%
Here qudits 0 and 1 hold by Alice and qudit 2 hold by Bob.

Applying the following two gates, we obtain%
\begin{eqnarray}
&&H_{0}C_{01}^{\dagger }\left\vert \phi \right\rangle _{0}\left\vert \Psi
^{\left( 0\right) }\right\rangle _{12}  \notag \\
&=&\frac{1}{\sqrt{d}}\overset{d-1}{\underset{j=0}{\sum }}\overset{d-1}{%
\underset{j^{\prime }=0}{\sum }}a_{j}\frac{1}{\sqrt{d}}\overset{d-1}{%
\underset{n=0}{\sum }}e^{i2\pi jn/d}\left\vert n\right\rangle _{0}\left\vert
j^{\prime }-j\right\rangle _{1}\left\vert j^{\prime }\right\rangle _{2}.
\label{3}
\end{eqnarray}%
Here $C_{01}^{\dagger }$ is the conjugate operator of $C_{01}$, and
explicitly $C_{01}^{\dagger }=\overset{d-1}{\underset{j,k=0}{\sum }}%
\left\vert j\right\rangle _{0}\left\langle j\right\vert \otimes \left\vert
k-j\right\rangle _{1}\left\langle k\right\vert $. For $d=2$ case, we have $%
C_{01}^{\dagger }=C_{01}$, however, $C_{01}^{\dagger }\neq C_{01}$ when $d>2$%
. By inserting identity $I_{1}=\overset{d-1}{\underset{m=0}{\sum }}%
\left\vert m\right\rangle _{1}\left\langle m\right\vert $ into Eq.(\ref{3})
and straightforward calculating, we have%
\begin{equation}
H_{0}C_{01}^{\dagger }\left\vert \phi \right\rangle _{0}\left\vert \Psi
^{\left( 0\right) }\right\rangle _{12}=\frac{1}{d}\overset{d-1}{\underset{n=0%
}{\sum }}\overset{d-1}{\underset{m=0}{\sum }}\left\vert n\right\rangle
_{0}\left\vert m\right\rangle _{1}X^{m}Z^{n}\left\vert \phi \right\rangle
_{2}.  \label{4}
\end{equation}%
Now it is interesting to note that when qudits 0 and 1 are in computational
basis $|n\rangle _{0},|m\rangle _{1}$, qudit 2 is the state $X^{m}Z^{n}|\phi
_{0}\rangle $. Apparently, by performing single-qudit measurement on qudits
0 and 1 in computational basis, Alice can get $d^{2}$ \ possible results
with equal probability $\frac{1}{d^{2}}$. According to Alice's measurement
result, Bob then applies corresponding operator $(X^{m}Z^{n})^{\dagger }$ on
qudit 2, Bob's final state is the original state $|\phi \rangle $. That is
to say, if Alice gets $\left\vert n\right\rangle _{0}\left\vert
m\right\rangle _{1}$ by measurement, Bob can recover the state $\left\vert
\phi \right\rangle _{2}$ on qudit 2 by applying corresponding unitary
transformation $(X^{m}Z^{n})^{\dagger }$. This is in principle the quantum
teleportation in $d$-dimensional Hilbert space. The only difference between
this scheme and the original one in Ref.\cite{teleportation} is that the
generalized Bell measurement in $d$-dimension is replaced by a generalized
CNOT and Hadamard operation followed by two single-qudit measurement
performed only by Alice, i.e., it is still a local operation. The classical
communication is certainly necessary to ensure that Bob can recover the
teleported state. We remark that when $d=2$, it reduces to the case of qubit
teleportation. This teleportation scheme is depicted in the Fig.1.

\begin{figure}[ht]
\includegraphics[width=6cm]{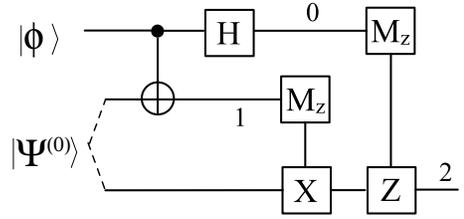} \vskip 1truecm
\caption{In the teleportation, the Bell measurement by Alice is replaced by
a CNOT gate and a Hadarmard gate. The measurement is performed on individual
qubits 0 and 1, and the measurement results is communicated through
classical channel. According to the measurement results, Bob applies
corresponding unitary transformations on 2 to recover the initial state. For
$d$-dimensional case, the procedures are almost the same, however, to
recover the initial state, the applied matrices on state 2 are the
generalized Pauli matrices $(X^mZ^n)^{\dagger }$ corresponding to the
measurement results $m,n$ on states $0$ and $1$.}
\end{figure}

Actually there are another scheme to realize single-qubit measurement
teleportation, which is similar to the method discussed in above. Here we
just give the direct calculating result as follows,%
\begin{equation}
H_{1}C_{10}^{\dagger }\left\vert \phi \right\rangle _{0}\left\vert \Psi
^{\left( 0\right) }\right\rangle _{12}=\frac{1}{d}\overset{d-1}{\underset{n=0%
}{\sum }}\overset{d-1}{\underset{m=0}{\sum }}\left\vert n\right\rangle
_{0}\left\vert m\right\rangle _{1}Z^{m}X^{-n}\left\vert \phi \right\rangle
_{2}.  \label{5}
\end{equation}%
In $d$-dimensional Hilbert space, $X^{-1}\left\vert j\right\rangle
=\left\vert \left( j-1\right) \text{mod}d\right\rangle $ and $X^{-1}\neq X$
when $d\neq 2$. The only difference between this scheme and the scheme in (%
\ref{4}) is that here the controlled qudit and the target qudit exchanged
with each other.

As we mentioned, in real physical systems, CNOT gate usually can only be
performed for nearby two particles. However, restricted to condition that
the particles interaction is only available for nearest neighboring sites,
if we have a prior maximally entangled state, CNOT gate for remotely
separated particles can be still realized. This remote CNOT for qubits case
has already been studied, for example in Ref.\cite{gottesman,H
Representation} and the remote controlled phase gate which is equivalent to
remote CNOT by local operation in Ref.\cite{DK}. However, the remote CNOT
for general $d$-dimensional states is not yet available. We will next
present the remote CNOT gate for $d$-dimensional case. For completeness,
let's first consider the remote CNOT in two-dimension.

In two-dimensional Hilbert space, arbitrary quantum state of two qubits,
which may be product state or coherent state, can be written as%
\begin{equation}
\left\vert \psi \right\rangle _{ij}=\alpha _{00}\left\vert 00\right\rangle
_{ij}+\alpha _{01}\left\vert 01\right\rangle _{ij}+\alpha _{10}\left\vert
10\right\rangle _{ij}+\alpha _{11}\left\vert 11\right\rangle _{ij},
\label{6}
\end{equation}%
where $\left\vert \alpha _{00}\right\vert ^{2}+\left\vert \alpha
_{01}\right\vert ^{2}+\left\vert \alpha _{10}\right\vert ^{2}+\left\vert
\alpha _{11}\right\vert ^{2}=1$. We have already known the CNOT operator $%
C_{ij}=\left\vert 0\right\rangle _{i}\left\langle 0\right\vert \otimes
I_{j}+\left\vert 1\right\rangle _{i}\left\langle 1\right\vert \otimes X_{j}$%
, and then after performing CNOT on state $\left\vert \psi \right\rangle
_{ij}$, we have%
\begin{equation}
C_{ij}\left\vert \psi \right\rangle _{ij}=\alpha _{00}\left\vert
00\right\rangle _{ij}+\alpha _{01}\left\vert 01\right\rangle _{ij}+\alpha
_{10}\left\vert 11\right\rangle _{ij}+\alpha _{11}\left\vert 10\right\rangle
_{ij}.  \label{7}
\end{equation}%
Here $i$ denotes control-qubit and $j$ denotes target-qubit.

Now suppose qubits $i$ and $j$ are located remotely, then we can not perform
CNOT between them directly. On the other hand, qubit $i^{\prime }$ is in the
nearest neighboring site with qubit $i$, similarly qubits $j,j^{\prime }$
are in nearest neighboring sites. Then CNOT gates can be performed directly
on pairs $(i,i^{\prime })$ and $(j,j^{\prime })$. And moreover, we suppose
that $i^{\prime }$ and $j^{\prime }$ share a maximally entangled state $%
\left\vert \Psi ^{\left( 0\right) }\right\rangle _{i\prime j\prime }=\frac{1%
}{\sqrt{2}}\left( \left\vert 00\right\rangle _{i\prime j\prime }+\left\vert
11\right\rangle _{i\prime j\prime }\right) $. Let's consider that states $%
i,i^{\prime }$ belong to Alice, and $j,j^{\prime }$ belong to Bob. Here in
order to see the role of teleportation in the remote CNOT, we divide the
scheme of remote CNOT gate into several steps. \textit{Step} 1, Bob applies
CNOT operator $C_{j\prime j}$ on pair $(j,j^{\prime })$locally, then
performs Hadamard gate operate on $j^{\prime }$. Thus we have%
\begin{eqnarray}
&&H_{j\prime }C_{j\prime j}\left( \left\vert \psi \right\rangle _{ij}\otimes
\left\vert \Psi ^{\left( 0\right) }\right\rangle _{i\prime j\prime }\right)
\notag \\
&=&\frac{1}{2}\left[ \left\vert \psi \right\rangle _{ii^{\prime }}\left\vert
00\right\rangle _{j\prime j}+X_{i^{\prime }}\left\vert \psi \right\rangle
_{ii^{\prime }}\left\vert 01\right\rangle _{j\prime j}\right.  \notag \\
&&\left. +Z_{i^{\prime }}\left\vert \psi \right\rangle _{ii^{\prime
}}\left\vert 10\right\rangle _{j\prime j}+Z_{i^{\prime }}X_{i^{\prime
}}\left\vert \psi \right\rangle _{ii^{\prime }}\left\vert 11\right\rangle
_{j\prime j}\right] .  \label{8}
\end{eqnarray}%
Obviously, this is the same as the teleportation operation as we presented,
which can teleport $\left\vert \psi \right\rangle _{ij}$ to $\left\vert \psi
\right\rangle _{ii^{\prime }}$. \textit{Step} 2, Alice applies CNOT operator
on her side $C_{ii^{\prime }}$. By these two steps, the initial joint state
becomes%
\begin{eqnarray}
&&C_{ii\prime }H_{j\prime }C_{j\prime j}\left( \left\vert \psi \right\rangle
_{ij}\otimes \left\vert \Psi ^{\left( 0\right) }\right\rangle _{i\prime
j\prime }\right)  \notag \\
&=&\frac{1}{2}\left[ C_{ij}\left\vert \psi \right\rangle _{ij}\otimes
\left\vert 00\right\rangle _{i\prime j\prime }+Z_{i}C_{ij}\left\vert \psi
\right\rangle _{ij}\otimes \left\vert 01\right\rangle _{i\prime j\prime
}\right.  \notag \\
&&\left. +X_{j}C_{ij}\left\vert \psi \right\rangle _{ij}\otimes \left\vert
10\right\rangle _{i\prime j\prime }-Z_{i}X_{j}C_{ij}\left\vert \psi
\right\rangle _{ij}\otimes \left\vert 11\right\rangle _{i\prime j\prime }
\right] .  \notag \\
&&  \label{9}
\end{eqnarray}%
Now we may notice that the states of $i,j$ in each term of the superposition
are always related with CNOT gate of $|\psi \rangle $. \textit{Step} 3,
Alice and Bob measures qubit $i^{\prime }$ and $j^{\prime }$ at their own
sites in computational basis respectively. By measurement, they can have
four results $\left\vert 00\right\rangle $, $\left\vert 01\right\rangle $, $%
\left\vert 10\right\rangle $ and $\left\vert 11\right\rangle $, each occurs
with probability $\frac{1}{4}$. \textit{Step} 4, by exchanging the classical
information of measurement results, Alice and Bob can apply corresponding
Pauli operators on qubit $i$ and $j$ respectively. Thus the CNOT gate $%
C_{ij} $ is realized between remotely separated qubit $i$ and qubit $j$.
This is the remote CNOT gate. This processing is depicted in Fig.2

\vskip 1truecm

\begin{figure}[ht]
\includegraphics[width=7cm]{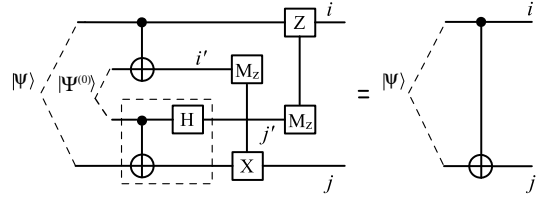} \vskip 1truecm
\caption{The remote CNOT gate is constituted by two CNOT gates which are
performed on nearby qubits or qudits and the Hadamard gates. After the
measurement in computational basis represented by $M_z$, Alice and Bob in
sites $i$ and $j$ apply corresponding transformation according to the
measurement results, the CNOT gate is finally realized for remotely
separated parties $i$ and $j$. The dashed square part in the scheme is the
same as that of the teleportation. Similarly, the $d$-dimensional remote
CNOT gate can be realized where the generalized Pauli matrices should be
applied according to the measurement results.}
\end{figure}

\vskip 1truecm

Now, we shall present our result of the $d$-dimensional remote CNOT gate. In
above, we have already defined some operators in $d$-dimensional Hilbert
space, we shall now use them here again. Arbitrary state of two qudits in $d$%
-dimensional Hilbert space can be expressed as%
\begin{equation}
\left\vert \psi \right\rangle _{ij}=\overset{d-1}{\underset{k,l=0}{\sum }}%
\alpha _{kl}\left\vert kl\right\rangle _{ij},  \label{10}
\end{equation}%
where $\overset{d-1}{\underset{k,l=0}{\sum }}\left\vert \alpha
_{kl}\right\vert ^{2}=1$. Also let $i$ be the control-qudit and $j$ the
target-qudit, the CNOT gate is represented as,
\begin{equation}
C_{ij}\left\vert \psi \right\rangle _{ij}=\overset{d-1}{\underset{k,l=0}{%
\sum }}\alpha _{kl}\left\vert k\left( l+k\right) \right\rangle _{ij}.
\label{11}
\end{equation}%
A prior maximally entangled state shared between Alice and Bob can be
written as%
\begin{equation}
\left\vert \Psi ^{\left( 0\right) }\right\rangle _{i^{\prime }j^{\prime }}=%
\frac{1}{\sqrt{d}}\overset{d-1}{\underset{m=0}{\sum }}\left\vert
mm\right\rangle _{i^{\prime }j^{\prime }}.  \label{12}
\end{equation}%
Following the same steps as in two-dimensional remote CNOT case, we obtain%
\begin{eqnarray}
&&C_{ii^{\prime }}H_{j^{\prime }}C_{j^{\prime }j}\left\vert \psi
\right\rangle _{ij}\otimes \left\vert \Psi ^{\left( 0\right) }\right\rangle
_{i^{\prime }j^{\prime }}  \notag \\
&=&\frac{1}{\sqrt{d}}\overset{d-1}{\underset{m=0}{\sum }}\overset{d-1}{%
\underset{k,l=0}{\sum }}\alpha _{kl}\left\vert k\left( l+m\right)
\right\rangle _{ij}\otimes \left\vert \left( m+k\right) \right\rangle
_{i^{\prime }}  \notag \\
&&\otimes \frac{1}{\sqrt{d}}\overset{d-1}{\underset{n=0}{\sum }}e^{\frac{%
2\pi inm}{d}}\left\vert n\right\rangle _{j^{\prime }}  \notag \\
&=&\frac{1}{d}\overset{d-1}{\underset{m=0}{\sum }}\overset{d-1}{\underset{n=0%
}{\sum }}e^{\frac{2\pi inm}{d}}Z_{i}^{-n}X_{j}^{m}C_{ij}^{\dagger
}\left\vert \psi \right\rangle _{ij}\otimes \left\vert mn\right\rangle
_{i^{\prime }j^{\prime }},  \label{13}
\end{eqnarray}%
where the generalized Hadamard gate $H=\frac{1}{\sqrt{d}}\overset{d-1}{%
\underset{j,n=0}{\sum }}e^{i2\pi jn/d}\left\vert n\right\rangle \left\langle
j\right\vert $ is equivalent to quantum Fourier transform, and $%
C_{ij}\left\vert kl\right\rangle _{ij}=\left\vert k\left( l+k\right)
\right\rangle _{ij}$, $C_{ij}^{\dagger }\left\vert kl\right\rangle
_{ij}=\left\vert k\left( l-k\right) \right\rangle _{ij}$, $X_{j}\left\vert
l\right\rangle _{j}=\left\vert \left( l+1\right) \text{mod}d\right\rangle
_{j}$, $Z_{i}\left\vert k\right\rangle _{i}=e^{i2\pi k/d}\left\vert
k\right\rangle _{i}$. Fortunately, still we may notice that in the
superposition, state of $i,j$ in each term is always the CNOT gate on $|\psi
\rangle $ though extra generalized Pauli matrices are present in front of
it. However, by measurement on sites $i^{\prime },j^{\prime }$, those Pauli
matrices can be deleted if we apply the corresponding transformations. Now
the scheme is like the following: Alice (Bob) measure qudit $i^{\prime }$ ($%
j^{\prime }$) at her (his) site in computational basis. They will know the
state $|m\rangle _{i^{\prime }},|n\rangle _{j^{\prime }}$, and each will
occur with probability $\frac{1}{d^{2}}$. Then Alice and Bob exchange the
classical information of measurement results and apply corresponding Pauli
operators on qudits $i$ and $j$ respectively. Thus the remote CNOT gate $%
C_{ij}^{\dagger }$\ between spatially separated qudits $i$ and $j$ is
realized. We note that in $d$-dimensional Hilbert space the final result
which is realized is actually the conjugate operator of remote CNOT gate,
i.e., $C_{ij}^{\dagger }=\overset{d-1}{\underset{k,l=0}{\sum }}\left\vert
k\right\rangle _{i}\left\langle k\right\vert \otimes \left\vert
l-k\right\rangle _{j}\left\langle l\right\vert $. This, however, can be
easily changed by a transformation, i.e., the CNOT and Hadarmad in l.h.s. of
(\ref{13}) are changed to their conjugate forms, and the conjugate CNOT in
r.h.s. is changed back to CNOT. The reason that we present this form is that
our generalization from two-dimensional case to $d$-dimensional case seems
more explicit. When $d=2$, $Z^{-1}=Z$ and $C_{ij}^{\dagger
}=C_{ij}=\left\vert 0\right\rangle _{i}\left\langle 0\right\vert \otimes
I_{j}+\left\vert 1\right\rangle _{i}\left\langle 1\right\vert \otimes X_{j}$%
, it recovers the Eq.(\ref{9}).

From above, we know that remote CNOT gate can be considered as a combination
of teleportation operation and a local CNOT operation. The opposite
direction is that a local CNOT can be teleported through a maximally
entangled state. For qubit case, the universal quantum gates teleported by
optical system has been proposed in Ref.\cite{quantum gate}. Finally we
would like to emphasize that the two local CNOT gates and the measurements
in the scheme of Fig.2 can be performed in parallel which can be realized at
the same time. This scheme is different from the remote CNOT scheme proposed
in Refs. \cite{PRA65,PRA71,ph05}. In their nonlocal operations a
state-operator which is named as "stator" need to be prepared first, then a
CNOT and a measurement can be performed. That means the two local CNOT and
measurements in their scheme should be performed in time order. In this
sense, the scheme presented in this paper has the advantage of parallel in
computation. Note that our scheme is also different from the scheme in Ref.%
\cite{remote CNOT}.

Restricted to condition that only nearest neighboring interaction is usually
available in real systems, CNOT gate can only be performed on nearby quantum
states. In case CNOT gate is necessary for remotely separated states, the
remote CNOT gate is an ideal choice where an extra maximally entangled state
is needed. By observing that the remote CNOT gate is actually a combination
of teleportation and a CNOT for qubit system, we successfully formulate the $%
d$-dimensional remote CNOT gate similarly by a $d$-dimensional teleportation
and a CNOT gate. In the scheme of remote CNOT gate, the teleportation by
Bell measurement is replaced by single state measurement scheme. Our result
can be used not only to realize CNOT remotely, but also it can be used to
teleport universal quantum gates. Compared with scheme in which states
transfer is realized by a series of swap gates, the remote CNOT scheme can
be achieved with the expense of one extra maximally entangled state. Thus it
can be considered to be an economy way.

\textit{Acknowledgements}: HF was supported by "Bairen" program, NSFC grant
(10674162) and "973" program (2006CB921107).

\end{document}